\documentclass[prb,aps,superscriptaddress,twocolumn,amssymb,showpacs]{revtex4}

\usepackage{color}
\usepackage{graphicx}
\usepackage{longtable}
\usepackage{epsfig}
\usepackage{dcolumn}
\usepackage{bm}
\usepackage{amssymb}

\begin{document}

\title{      Hidden dimer order in the quantum compass model}

\author{     Wojciech Brzezicki}
\affiliation{Marian Smoluchowski Institute of Physics, Jagellonian
             University, Reymonta 4, PL-30059 Krak\'ow, Poland }

\author {    Andrzej M. Ole\'{s} }
\affiliation{Marian Smoluchowski Institute of Physics, Jagellonian
             University, Reymonta 4, PL-30059 Krak\'ow, Poland }
\affiliation{Max-Planck-Institut f\"ur Festk\"orperforschung,
             Heisenbergstrasse 1, D-70569 Stuttgart, Germany  }

\date{4 June 2010}

\begin{abstract}
We introduce an exact spin transformation that maps frustrated
$Z_{i,j}Z_{i,j+1}$ and $X_{i,j}X_{i+1,j}$ spin interactions along
the rows and columns of the quantum compass model (QCM) on an
$L\times L$ square lattice to $(L-1)\times (L-1)$ quantum spin
models with $2(L-1)$ classical spins. Using the symmetry
properties we unravel the hidden dimer order in the QCM, with
equal two-dimer correlations $\langle
X_{i,i}X_{i+1,i}X_{k,l}X_{k+1,l}\rangle$ and $\langle
X_{i,i}X_{i+1,i}X_{l,k}X_{l+1,k}\rangle$ in the ground state,
which is independent of the actual interactions. This order
coexists with Ising-like spin correlations which decay with
distance.

{\it Published in: Phys. Rev. B {\bf 82}, 060401 (2010).}
\end{abstract}


\maketitle

The quantum compass model (QCM) originates from the frustrated
(Kugel-Khomskii) superexchange \cite{Kho82} in transition metal
oxides with degenerate $3d$ orbitals. Recent interest in this
model is motivated by its interdisciplinary character as it plays
a role in the variety of phenomena beyond the correlated oxides.
It describes a quantum phase transition between competing types of
order when anisotropic interactions are varied through the
isotropic point, as shown by an analytical method, \cite{Dou05}
mean-field (MF), \cite{Che07} and numerical studies.
\cite{Mil05,Wen08,Oru09,Brz10,Tro10} The QCM is dual to the models
of $p+ip$ superconducting arrays \cite{Nus05} and to the toric
code model in a transverse field. \cite{Vid09} It was also
suggested as an effective model for Josephson arrays of protected
qubits, \cite{Dou05} as realized in recent experiment.
\cite{Gla09} Finally, it could describe polar molecules in optical
lattices and systems of trapped ions. \cite{Mil07}

In spite of several numerical studies,
\cite{Mil05,Wen08,Oru09,Brz10,Tro10} the nature of spin order in
the two-dimensional (2D) QCM is not yet fully understood. By an
exact solution of the QCM on a ladder we have shown, however, that
the invariant subspaces may be deduced using the symmetry.
\cite{Brz09} The 2D QCM shows a self-duality \cite{Nus05} which
might serve to reveal nontrivial hidden symmetries. \cite{Cob10}
In this Letter we employ exact spin transformations which allow us
to discover a surprising hidden dimer order in the QCM which
manifest itself by exact relations between four-point correlation
functions in the ground state. We also demonstrate nonlocal MF
splitting of the QCM in the ground subspace and determine spatial
decay of spin correlations in the thermodynamic limit.

{\it Reduced Hamiltonian.} --- We consider the anisotropic
ferromagnetic QCM for pseudospins 1/2 on an $L\times L$ square
lattice with periodic boundary conditions (PBC),
\begin{equation}
{\cal H}(\alpha)=-\sum_{i,j=1}^{L} \left\{ (1-\alpha)
X_{i,j}X_{i+1,j} + \alpha Z_{i,j}Z_{i,j+1} \right\}, \label{ham}
\end{equation}
where $\{X_{i,j},Z_{i,j}\}$ stand for Pauli matrices at site
$(i,j)$, i.e., $X_{i,j}\equiv\sigma^{x}_{i,j}$ and
$Z_{i,j}\equiv\sigma^{z}_{i,j}$ components, interacting on
vertical and horizontal bonds. In case of $L$ being even, this
model is equivalent to the antiferromagnetic QCM. We can easily
construct a set of $2L$ operators which commute with the
Hamiltonian but anticommute with one another:\cite{Dou05}
$P_i\equiv\prod_{j=1}^L\,X_{i,j}$ and
$Q_j\equiv\prod_{i=1}^L\,Z_{i,j}$. Below we will use as symmetry
operations all $R_i\equiv P_i P_{i+1}$ and $Q_j$ to reduce the
Hilbert space; this approach led to the exact solution of the
compass ladder. \cite{Brz09} The QCM Eq. (1) can be written in
common eigenbasis of $\{R_i,Q_j\}$ operators using:
\begin{eqnarray}
\label{transx} X_{i,j}&=&\prod_{p=i}^L \tilde {X}_{p,j}\,, \hskip
1cm \tilde {X}_{i,j}=X'_{i,j-1} X'_{i,j}\,,  \\
\label{transz} Z_{i,j}&=&\tilde {Z}_{i-1,j} \tilde {Z}_{i,j}\,,
\hskip .7cm \tilde {Z}_{i,j}=\prod_{q=j}^L Z'_{i,q}\,,
\end{eqnarray}
where $\tilde {Z}_{0,j}\equiv 1$ and $X'_{i,0}\equiv 1$. One finds
that the transformed Hamiltonian, ${\cal
H}'(\alpha)=-(1-\alpha)H'_x-\alpha H'_z$, contains no
$\tilde{X}_{L,j}$ and no $Z'_{i,L}$ operators so the corresponding
$\tilde {Z}_{L,j}$ and $X'_{i,L}$ can be replaced by their
eigenvalues $q_j$ and $r_i$, respectively. The Hamiltonian ${\cal
H}'(\alpha)$ is dual to the QCM ${\cal H}(\alpha)$ of Eq. (1) in
the thermodynamic limit; we give here an explicit form of its
$x$-part,
\begin{eqnarray}
H'_x &=&  \sum_{i=1}^{L-1} \left\{ \sum_{j=1}^{L-2}X'_{i,j}
X'_{i,j+1} + X'_{i,1}+r_i X'_{i,L-1}\right\} \nonumber \\
&+&P'_{1} +\sum_{j=1}^{L-2}P'_j P'_{j+1}+r P'_{L-1} ,
\label{Hiks}
\end{eqnarray}
where $r=\prod_{i=1}^{L-1}r_i$, and new nonlocal
$P'_j=\prod_{p=1}^{L-1}\,X'_{p,j}$ operators originate from the
PBC. The $z$-part $H'_z$ follows from $H'_x$ by lattice
transposition $X'_{i,j}\rightarrow Z'_{i,j}$, and by
$r_i\rightarrow s_j=q_j q_{j+1}$. Ising variables $r_i$ and $s_j$
are the eigenvalues of the symmetry operators $R_i$ and
$S_j=Q_jQ_{j+1}$.

Instead of the initial $L\times L$ lattice of quantum spins, one
finds here $(L-1)\times(L-1)$ internal quantum spins with $2(L-1)$
classical boundary spins. The missing spin is related to the $Z_2$
symmetry of the QCM and makes every energy level at least doubly
degenerate. Although the form of Eq. (\ref{Hiks}) is complex, the
size of the Hilbert space is reduced in a dramatic way by a factor
$2^{2L-1}$ \cite{Brz10} which makes it possible to perform easily
exact (Lanczos) diagonalization of 2D $L\times L$ clusters up to
$L=5$ ($L=6)$.

{\it Equivalent subspaces.} --- The original QCM of Eq. (1) is
invariant under the transformation $X'\leftrightarrow Z'$, if one
also transforms the interactions, $\alpha\leftrightarrow
(1-\alpha)$. This implies that subspaces $(\vec{r},\vec{s})$ and
$(\vec{s},\vec{r})$ give the same energy spectrum which sets an
equivalence relation between the subspaces --- two subspaces are
equivalent means that the QCM (1) has in them the same energy
spectrum. This relation becomes especially simple for
$\alpha=\frac{1}{2}$ when for all $r_i$'s and $s_i$'s subspaces
$(\vec{r},\vec{s})$ and $(\vec{s},\vec{r})$ are equivalent.

Now we will explore another important symmetry of the 2D compass
model reducing the number of nonequivalent subspaces --- the
translational symmetry. We note from Eq. (\ref{Hiks}) that the
reduced Hamiltonians are not translationally invariant for any
choice of $(\vec{r},\vec{s})$ even though the original Hamiltonian
is. This means that translational symmetry must impose some
equivalence conditions among subspace labels
$\{\vec{r},\vec{s}\}$. To derive them, let's focus on translation
along the rows of the lattice by one lattice constant. Such
translation does not affect the $P_i$ symmetry operators, because
they consist of spin operators multiplied along the rows, but
changes $Q_j$ into $Q_{j+1}$ for all $j<L$ and $Q_L\rightarrow
Q_{1}$. This implies that two subspaces
$(\vec{r},q_1,q_2,\dots,q_L)$ and
$(\vec{r},q_L,q_1,q_2,\dots,q_{L-1})$ are equivalent for all
values of $\vec{r}$ and $\vec{q}$.
Now this result must be translated into the language of
$(\vec{r},\vec{s})$ labels, with $s_j=q_j q_{j+1}$ for all $j<L$.
This is two-to-one mapping because for any $\vec{s}$ one has two
$\vec{q}$'s such that $\vec{q}_+=(1,s_1,s_1s_2,\dots,s_1s_2\dots
s_{L-1})$ and $\vec{q}_-=-\vec{q}_+$ differ by global inversion.
This sets additional equivalence condition for subspace labels
$(\vec{r},\vec{s})$: two subspaces $(\vec{r},\vec{u})$ and
$(\vec{r},\vec{v})$ are equivalent if two strings
$(1,u_1,u_1u_2,\dots,u_1u_2\dots u_{L-1})$ and
$(1,v_1,v_1v_2,\dots,v_1v_2\dots v_{L-1})$ are related by
translations or by a global inversion. For convenience let us call
these two vectors TI (translation-inversion) related. Lattice
translations along the columns set the same equivalence condition
for $\vec{r}$ labels. Thus full equivalence conditions for
subspace labels of the QCM are:
\begin{itemize}
\item For $\alpha=\frac{1}{2}$ two subspaces $(\vec{r},\vec{s})$
and $(\vec{u},\vec{v})$ are equivalent if $\vec{r}$ is TI-related
with $\vec{u}$ and $\vec{s}$ with $\vec{v}$ or if $\vec{r}$ is
TI-related with $\vec{v}$ and $\vec{s}$ with $\vec{u}$.
\item For $\alpha\not=\frac{1}{2}$ two subspaces
$(\vec{r},\vec{s})$ and $(\vec{u},\vec{v})$ are equivalent if
$\vec{r}$ is TI-related with $\vec{u}$ and $\vec{s}$ with
$\vec{v}$.
\end{itemize}
We have verified that no other equivalence conditions exist
between the subspaces by numerical Lanczos diagonalizations for
lattices of sizes up to $6\times 6$, so we can change all {\it
if\/} statements above into {\it if and only if\/} ones.

\begin{figure}[t!]
\includegraphics[width=8.5 cm]{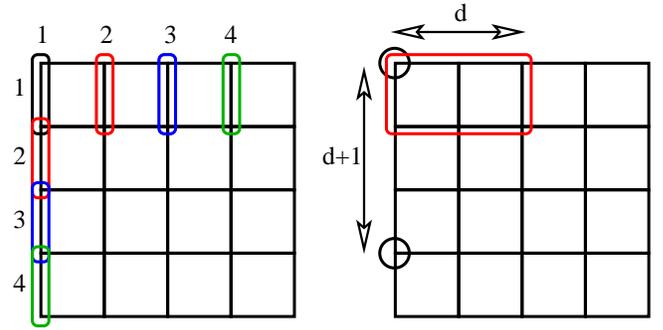}
\caption{(color online).
Example of the application of the proved
identities in two cases: left panel --- Eq. (\ref{fin}) for a
chosen dimer ${\cal D}_{i,j}\equiv X_{i,j}X_{i+1,j}$ (label 1),
correlations of ${\cal D}_{i,j}$ with a given ${\cal D}_{k,l}$ are 
the same for dimers ${\cal D}_{k,l}$ marked with the same label 
(color); right panel --- Eq. (\ref{mult}) long range correlation 
function $\langle X_{i,j}X_{i+d+1,j}\rangle$ along the column 
(circles) is equal to the $2d$--point $\langle XX\dots X\rangle$ 
correlation function along the row (frame of length $d$).} 
\label{fig:appl}
\end{figure}

{\it Hidden dimer order.} --- Due to the symmetries of the QCM Eq.
(1) only $\langle Z_{i,j}Z_{i,j+d}\rangle$ and $\langle
X_{i,j}X_{i+d,j}\rangle$ spin correlations are finite ($d>0$).
This suggests
that the entire spin order concerns {\it pairs of spins\/} from
one row (column) which could be characterized by four--point
correlation functions of the dimer-dimer type. Indeed, examining
such quantities for finite QCM clusters via Lanczos diagonalization 
we observed certain surprising symmetry: for any $\alpha$ 
dimer-dimer correlators $\langle{\cal D}_{i,j}{\cal D}_{k,l}\rangle$,
with ${\cal D}_{i,j}\equiv X_{i,j}X_{i+1,j}$,
are invariant under the reflection of the second dimer with 
respect to the diagonal passing through site $(i,j)$,
see left panel of Fig. \ref{fig:appl}. This general relation between 
correlation functions of the QCM will be proved below.

We will prove that in the ground state of the QCM for any two
sites $(i,j)$ and $(k,l)$ and for any $0<\alpha<1$:
\begin{eqnarray}
&& \langle X_{i,j}X_{i+1,j}X_{k,l}X_{k+1,l}\rangle \nonumber \\
&\equiv&
\langle X_{i,j}X_{i+1,j}X_{l-\delta,k+\delta}X_{l-\delta+1,k+\delta}\rangle ,
\label{thes}
\end{eqnarray}
where $\delta=j-i$,\cite{notezz} i.e., the second dimer is
reflected with respect to the diagonal. To prove Eq. (\ref{thes}) let us
transform again the effective Hamiltonian (\ref{Hiks}) in the ground
subspace ($r_i\equiv s_i\equiv 1$) introducing new spin operators
\begin{equation}
Z'_{i,j}=\tilde{Z}_{i,j}\tilde{Z}_{i,j+1} , \hskip 1cm
X'_{i,j}=\prod_{r=1}^j\tilde{X}_{i,r} ,
\label{trans2}
\end{equation}
with $i,j=1,\dots,L-1$ and $\tilde{Z}_{i,L}\equiv 1$. This yields to
\begin{eqnarray}
\label{Hiks2} \tilde{H}_x &=&  \sum_{i=1}^{L-1}
\sum_{j=1}^{L-1}\tilde{X}_{i,j}
+\prod_{i=1}^{L-1} \prod_{j=1}^{L-1}\tilde{X}_{i,j} \nonumber \\
&+& \sum_{i=1}^{L-1} \prod_{j=1}^{L-1}\tilde{X}_{i,j} +
    \sum_{i=1}^{L-1} \prod_{j=1}^{L-1}\tilde{X}_{j,i} , \\
\tilde{H}_z &=& \sum_{a}\left\{\sum_{b}\tilde{Z}_{a,b} +
\sum_{i=1}^{L-2}\left(\tilde{Z}_{a,i}\tilde{Z}_{a,i+1} +
\tilde{Z}_{i,a}\tilde{Z}_{i+1,a}\right)\right\}\nonumber \\
&+&\sum_{i=1}^{L-2} \sum_{j=1}^{L-2}
\tilde{Z}_{i,j}\tilde{Z}_{i,j+1}\tilde{Z}_{i+1,j}\tilde{Z}_{i+1,j+1} ,
\label{Hzet2}
\end{eqnarray}
where $a,b=1,L-1$. Due to the spin transformations
(\ref{transx},\ref{transz},\ref{trans2}), $\tilde{X}_{i,j}$
operators are related to the original bond operators by
$X_{i,j}X_{i+1,j}=\tilde{X}_{i,j}$, which implies that
\begin{equation}
\langle X_{i,j}X_{i+1,j}X_{k,l}X_{k+1,l} \rangle
= \langle \tilde{X}_{i,j}\tilde{X}_{k,l} \rangle .
\end{equation}

\begin{figure}[t!]
\includegraphics[width=5.5cm]{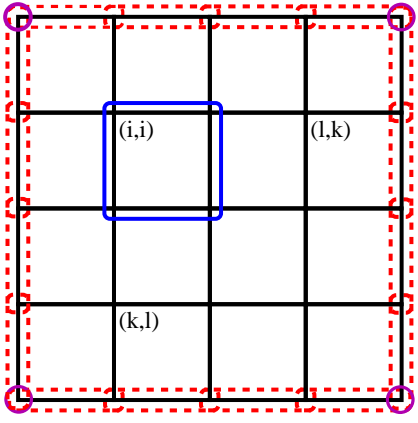}
\caption{(color online). Schematic view of the $z$-part of the
reduced ground subspace Hamiltonian (\ref{Hzet2}): circles in the
corners stand for $\tilde{Z}_{i,j}$ spin operators related to the
site $(i,j)$, dashed (red) frames are $\tilde{Z}\tilde{Z}$ operator
products acting on the boundaries of the lattice, and solid (blue)
square stands for one of the plaquette
$\tilde{Z}\tilde{Z}\tilde{Z}\tilde{Z}$ spin operators. The exemplary
three sites in the identity (\ref{fin}) are: $(i,i)$, $(k,l)$ and
$(l,k)$.}
\label{fig:hz}
\end{figure}

Because of the PBC, all original $X_{i,j}$ spins are equivalent,
so we choose $i=j$. The $x$-part (\ref{Hiks2}) of the Hamiltonian
is completely isotropic. Note that the $z$-part (\ref{Hzet2})
would also be isotropic without the boundary terms (see Fig.
\ref{fig:hz}); the effective Hamiltonian in the ground subspace
has the symmetry of a square. Knowing that in the ground state
we have only $Z_2$ degeneracy, one finds
\begin{equation}
\langle \tilde{X}_{i,i}\tilde{X}_{k,l} \rangle \equiv  \langle
\tilde{X}_{i,i}\tilde{X}_{l,k} \rangle, \label{fin}
\end{equation}
for any $i$ and $(k,l)$. This proves the identity (\ref{thes}) for
$\delta=0$; $\delta\neq 0$ case follows from lattice translations
along rows.

\begin{figure}[t!]
\includegraphics[width=7.5cm]{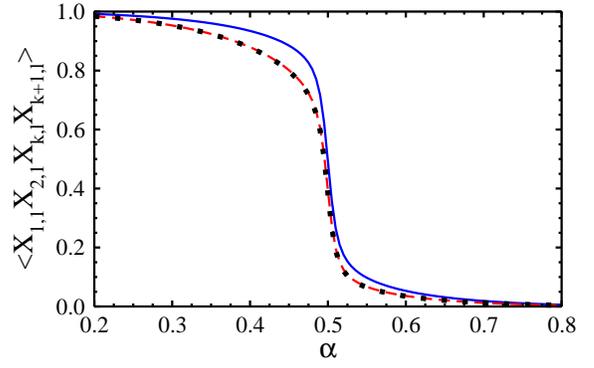}
\caption{(color online). Two-dimer $\langle
X_{1,1}X_{2,1}X_{k,l}X_{k+1,l}\rangle$ correlations for
$L=6$ and $0.2<\alpha<0.8$: $(k,l)=(1,2)$, $(1,3)$ and $(1,4)$
are shown by solid, dashed and dotted line. }
\label{fig:didi}
\end{figure}

The nontrivial consequences of Eq. (\ref{fin}) are: (i) {\it
hidden dimer order\/} in the ground state of the QCM, i.e., 
an "isotropic" behavior of the two-pair correlator in spite of
anisotropic interactions in the entire range of $0<\alpha<1$ 
(see Fig. \ref{fig:didi}), and 
(ii) long range two-site $\langle X_{i,j}X_{i+d+1,j}\rangle$ 
correlations of range $d$ along the columns which are equal to the 
multi-site $\langle XX\dots X\rangle$ correlations involving two 
neighboring rows, see right panel of Fig. \ref{fig:appl}. The latter 
follows from the symmetry properties of the transformed Hamiltonian 
(\ref{Hiks2},\ref{Hzet2}) applied to the multi-site correlations:
\begin{equation}
 \langle\tilde{X}_{i,i}\tilde{X}_{i,i+1}\dots\tilde{X}_{i,i+d}\rangle
=\langle\tilde{X}_{i,i}\tilde{X}_{i+1,i}\dots\tilde{X}_{i+d,i}\rangle.
\label{mult}
\end{equation}

{\it Mean-field approximation.} --- The $x$-part of the Hamiltonian
obtained from Eq. (\ref{Hiks}) in case of open boundaries reads:
\begin{equation}
H'_x = \sum_{i=1}^{L-1} \left\{ \sum_{j=1}^{L-2}X'_{i,j}
X'_{i,j+1} + X'_{i,1}+r_i X'_{i,L-1}\right\},
\end{equation}
and similarly for the $z$-part. In the ground subspace ($r_i\equiv
1$) this resembles the original QCM Eq. (1) but with linear
boundary terms, which should not affect the ground state
properties in the thermodynamic limit and can be regarded as
symmetry breaking fields, resulting in finite values of $\langle
X'_{i,j}\rangle$ and $\langle Z'_{i,j}\rangle$. Omitting the
boundary terms in $H'_x$ and $H'_z$ and putting infinite $L$ we
recover the 2D QCM written in nonlocal primed spin operators. Now
we can construct a MF splitting of the 2D lattice into
(ferromagnetic) Ising chains in transverse field, taking $\langle
Z'\rangle\equiv\langle Z'_{i,j}\rangle$ as a Weiss field 
for each row $i$:
\begin{equation}
{\cal H}'_i( \alpha) = -\sum_j\left\{(1-\alpha) X'_{i,j}X'_{i,j+1}
+ 2\alpha\langle Z'\rangle Z'_{i,j} \right\}.
\end{equation}
In analogy to the compass ladder, \cite{Brz09} it can be solved by
Jordan-Wigner transformation for each $i$:
\begin{eqnarray}
Z'_{i,j}&=&1-2c^{\dagger}_{i,j}c^{}_{i,j},  \\
X'_{i,j}&=&\left(c^{\dagger}_{i,j} e^{-i\frac{\pi}{4}}
                +c^{}_{i,j}e^{i\frac{\pi}{4}}\right)
\prod_{r<j}(1-2c^{\dagger}_{i,r}c^{}_{i,r}),
\end{eqnarray}
introducing fermion operators $\{c^{\dagger}_{i,j}\}$. The
diagonalization of the free fermion Hamiltonian can be completed
by performing first a Fourier transformation (from $\{j\}$ to
$\{k\}$) and next a Bogoliubov transformation (for $k>0$):
$\gamma^{\dagger}_{k}=\alpha^+_k c^{\dagger}_k+\beta^+_k
c^{}_{-k}$ and $\gamma^{\dagger}_{-k}=\alpha^-_k
c^{\dagger}_k+\beta^-_k c^{}_{-k}$, where
$\{\alpha^{\pm}_k,\beta^{\pm}_k\}$ are eigenmodes of the
Bogoliubov-de Gennes equation for the eigenvalues $\pm E_k$ (with
$E_k>0$). The resulting ground state is a vacuum of
$\gamma^{\dagger}_{k}$ fermion operators:
$|\Phi_0\rangle=\prod_{k>0}(\alpha^+_k+\beta^+_k
c^{\dagger}_{-k}c^{\dagger}_k)|0\rangle$, which can serve to
calculate correlations and the order parameter of the QCM in the
MF approach. In agreement with numerical results (not shown), the
only nonzero long range two-site spin correlation functions are:
$\langle X_{i,j}X_{i+d,j}\rangle$ and $\langle
Z_{i,j}Z_{i,j+d}\rangle$. For $d>1$ they can be represented as
follows:
\begin{eqnarray}
 \label{corelx}
\langle X_{i,j}X_{i+d,j} \rangle &=&
 \langle X'_{i,j}X'_{i,j+1} \rangle^d,  \\
 \label{corelz}
\langle Z_{i,j}Z_{i,j+d} \rangle &=&
 \langle Z'_{i,j}Z'_{i,j+1}\dots Z'_{i,j+d-1} \rangle ^2.
\end{eqnarray}

\begin{figure}[t!]
\includegraphics[width=7.0cm]{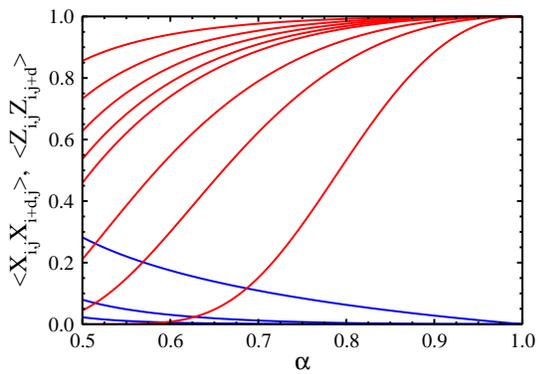}
\caption{(color online). Long range spin correlations of the 2D
QCM Eq. (1) obtained in the MF approach for $\alpha\ge\frac12$.
Lines starting from 1 at $\alpha=1$ are the $\langle
Z_{i,j}Z_{i,j+d}\rangle$ correlations (\ref{corelz}) for
$d=1,2,3,4,5,10,20,80$, while lines starting from 0 at $\alpha=1$
are the $\langle X_{i,j}X_{i+d,j}\rangle$ correlations
(\ref{corelx}) with $d=1,2,3$; both in descending order.}
\label{fig:cor}
\end{figure}

Having solved the self-consistency equation for $\langle
Z'\rangle=(1-2\langle n\rangle)$, with
$\langle n\rangle=
\frac{1}{L}\sum_{k>0}\left(\alpha^{-2}_k+\beta^{+2}_k\right)$,
one can easily obtain $\langle X'_{i,j}X'_{i,j+1}\rangle$
(\ref{corelx}) for increasing $\alpha$:
\begin{eqnarray}
\langle X'_{i,j}X'_{i,j+1} \rangle\!&=& \frac{2}{L}\sum_{k>0}
\left\{\cos{k}\,(\alpha^{-2}_k+\beta^{+2}_k)\right. \nonumber   \\
& &\left.\hskip .7cm
+\sin{k}\,(\alpha^-_k\beta^-_k-\alpha^+_k\beta^+_k)\right\}.
\end{eqnarray}
The nonlocal $\langle Z'_{i,j}Z'\dots Z'_{i,j+d-1}\rangle$
correlations (\ref{corelz}) are more difficult to find but they
can be approximated by
\begin{equation}
\langle Z'_{i,j}Z'_{i,j+1}\dots Z'_{i,j+d-1} \rangle=\prod_{k>0}
\left\{\!\alpha^{+2}_k\left(1-2\frac{d}{L}\right)^2\!+\beta^{+2}_k\!\right\},
\end{equation}
where $L\to\infty$ and $k=(2l-1)\frac{\pi}{L}$ with
$l=1,2,\dots,\frac{L}{2}$. This approximation is valid as long as
$d\ll L$. One finds that the long range $\langle
Z_{i,j}Z_{i,j+d}\rangle$ correlations in $Z$-ordered phase at
$\alpha\ge\frac12$ show the absence of the Ising-like long range
order for $\alpha<1$ (Fig. \ref{fig:cor}) --- they decrease slowly
with growing distance $d$ or decreasing $\alpha$. \cite{noteis} In
contrast, the $\langle X_{i,j}X_{i+d,j}\rangle$ correlations are
significant only for nearest neighbors ($d=1$) and close to
$\alpha=\frac{1}{2}$.

The advantage of this nonlocal MF approach for the QCM Eq. (1)
over the standard one, which takes $\langle Z\rangle$ as a Weiss
field, is that we do not break the $\{P_i,Q_j\}$ and $Z_2$
symmetries of the model. What more, thanks to numerical and
analytical results we know that order parameter of the QCM is
given by $\langle H_z\rangle$ \cite{Mil05} --- the quantity
behaving more like $\langle Z'\rangle$ rather than $\langle
Z\rangle$ (having $\langle Z\rangle>0$ would mean long range
magnetic order !). Another interesting feature of the Hamiltonian
(\ref{ham}) is that it describes all nonlocal compass excitations
over the ground state, while the local ones manifest themselves by
directions of symmetry breaking fields. These nonlocal column
(row) flips are especially interesting from the point of view of
topological quantum computing \cite{Dou05} because they guarantee
that the system is protected against local perturbations.

{\it Conclusions.} --- On the example of the QCM, we argue that
the properties of spin models which are not SU(2) symmetric can be
uniquely determined by discrete symmetries like parity. In this
case conservation of spin parities in rows and columns, for $x$
and $z$-components of spins, makes the system in the ground state
behave according to a nonlocal Hamiltonian (\ref{Hiks}).
\cite{notepa} In the ground state most of the two-site spin
correlations vanish and the two-dimer correlations exhibit the
nontrivial hidden order. The excitations involve whole lines of
spins in the lattice and occur in invariant subspaces which can be
classified by lattice translations --- the reduction of the
Hilbert space achieved in this way is important for future
numerical studies of the QCM and will play a role for spin models
with similar symmetries. Finally, the nonlocal Hamiltonian
containing symmetry breaking terms suggests the MF splitting
respecting conservation of parity and leading to the known physics
of one-dimensional quantum Ising model describing correlation
functions and the order parameter of the QCM.

We thank L. F. Feiner, P. Horsch and K. Ro\'sciszewski for
insightful discussions. We acknowledge support by the Foundation
for Polish Science (FNP) and by the Polish government under
project N202 069639.


\begin{thebibliography}{99}

\bibitem{Kho82} K. I. Kugel and D. I. Khomskii,
                   Sov. Phys. Usp. \textbf{25}, 231 (1982);
                D. I.~Khomskii and M. V.~Mostovoy,
                   J. Phys. A \textbf{36}, 9197 (2003);
                Z. Nussinov, M. Biskup, L. Chayes,
                   and J. van den Brink,
                   Europhys. Lett. \textbf{67}, 990 (2004).

\bibitem{Dou05} B.~Dou\c{c}ot, M. V. Feigel'man, L. B. Ioffe,
                   and A. S. Ioselevich,
                   \prb \textbf{71}, 024505 (2005).

\bibitem{Che07} H.-D. Chen, C. Fang, J. Hu, and H. Yao,
                   \prb \textbf{75}, 144401 (2007).

\bibitem{Mil05} J.~Dorier, F.~Becca, and F.~Mila,
                   \prb \textbf{72}, 024448 (2005).

\bibitem{Wen08} S.~Wenzel and W.~Janke,
                   \prb \textbf{78}, 064402 (2008).

\bibitem{Oru09} R. Or\'us, A. C. Doherty, and G. Vidal,
                   \prl \textbf{102}, 077203 (2009).

\bibitem{Brz10} W.~Brzezicki and A. M.~Ole\'s,
                   J. Phys.: Conf. Series \textbf{200}, 012017 (2010).

\bibitem{Tro10} L. Cincio, J. Dziarmaga, and A. M. Ole\'s,
                   \prb \textbf{82}, in press (2010), also arXiv:1001.5457;
                F. Trousselet, A. M. Ole\'s, and P. Horsch,
                   EPL \textbf{91}, in press (2010), also arXiv:1005.1508.

\bibitem{Nus05} C.~Xu and J. E.~Moore,
                   \prl \textbf{93}, 047003 (2004);
                Z.~Nussinov and E.~Fradkin,
                   \prb \textbf{71}, 195120 (2005).

\bibitem{Vid09} J.~Vidal, R. Thomale, K.~P.~Schmidt, and S.~Dusuel,
                  \prb \textbf{80}, 081104 (2009).

\bibitem{Gla09} S. Gladchenko {\it et al.\/},
                   Nature Phys. \textbf{5}, 48 (2009).

\bibitem{Mil07} P. Milman {\it et al.\/},
                   \prl \textbf{99}, 020503 (2007).

\bibitem{Brz09} W.~Brzezicki and A. M.~Ole\'s,
                   \prb \textbf{80}, 014405 (2009).

\bibitem{Cob10} E.~Cobanera, G.~Ortiz, and Z.~Nussinov,
                   \prl \textbf{104}, 020402 (2010).

\bibitem{notezz} The symmetry of Eq. (1) implies a similar relation
                 for $Z$-correlations with dimers along the horizontal
                 bonds.

\bibitem{noteis} At $\alpha=\frac12$ the $Z$-ordered and $X$-ordered
                 Ising-like phases of the QCM are degenerate, see
                 e.g. Ref. \onlinecite{Wen08}.

\bibitem{notepa} The nonlocal Hamiltonian is just the original QCM
                 Eq. (1) written in eigenbasis of $\{P_i,Q_j\}$
                 parities.

\end{thebibliography}
\end{document}